\documentclass{sig-alternate-05-2015}%
\usepackage{amsfonts}
\usepackage{amsmath}
\usepackage{amssymb}
\usepackage{graphicx}%
\usepackage{color, colortbl}
\usepackage{subfig}
\usepackage{float}
\usepackage{booktabs}
\usepackage{multirow}
\usepackage[inline]{enumitem}
\usepackage{tabularx}
\usepackage{todonotes}
\definecolor{Gray}{gray}{0.9}

\begin{document}


\title{PLIERS: a Popularity-Based Recommender System for Content Dissemination in Online Social Networks}
\author{Valerio Arnaboldi$^{1}$, Mattia G. Campana$^{1,2}$, Franca Delmastro$^{1}$,\\and Elena Pagani$^{2,1}$\\$^{1}$IIT-CNR - Via G. Moruzzi 1, 56124, Pisa, ITALY\\$^{2}$Computer Science Department, University of Milano, Milano, ITALY\\\{v.arnaboldi, m.campana, f.delmastro\}@iit.cnr.it, pagani@di.unimi.it}
\maketitle


\section{Introduction}

In this paper, we present \textit{PLIERS (PopuLarity-based ItEm Recommender System)}, a novel \emph{Tag-based Recommender systems} ({\sc tbrs}s)~\cite{ricci2011recommender} based on \emph{folksonomies}~\cite{Quattrone2012Measuring}. 
It relies on the assumption that a user is mainly interested in items and tags with popularity similar to that of the items she already owns, and that the similarity between items/tags can also highlight a semantic relationship between them. 
To evaluate PLIERS, we performed a set of experiments on real OSN datasets, demonstrating that it outperforms state-of-the-art solutions (described in Section \ref{sec:related2}) in terms of personalization, relevance, and novelty of recommendations by better describing the human behavior in  selecting new interesting contents.

\section{Notation and related work}
\label{sec:related2}
Formally, a folksonomy can be represented with three node sets: users $U = \{u_1, \ldots , u_n\}$, items $I = \{i_1, \ldots, i_m\}$ and tags $T = \{t_1, \ldots, t_k\}$.
Each binary relation between them can be described using adjacency matrices, \textbf{$A^{UI}$}, \textbf{$A^{IT}$}, \textbf{$A^{UT}$} respectively for user-item, item-tag and user-tag relations. If the user $u_l$ has collected the item $i_s$, we set $a^{UI}_{l,s} = 1$, $a^{UI}_{l,s} = 0$ otherwise. Similarly, $a^{IT}_{s,q} = 1$ if $i_s$ is tagged with $t_q$ and $a^{IT}_{s,q} = 0$ otherwise. Furthermore,  $a^{UT}_{l,q} = 1$ if $u_l$ owns items tagged with $t_q$, and $a^{UT}_{l,q} = 0$ otherwise. The three matrices can be represented as a tripartite graph $G^T=(U,I,T,E)$ where $U$, $I$, and $T$ are set of nodes representing users, items, and tags respectively, and $E$ is the set of edges between nodes corresponding to the elements equal to 1 in the matrices.
A bipartite graph $G^B=(U,V,E)$ may be used instead of a tripartite graph, with $U$ the set of users, and $V$ the set of either items or tags. In the following, we will consider bipartite user-item graphs with $n$ users and $m$ items where an edge between the user $u_l$ and the item $i_s$ indicates that $u_l$ owns $i_s$.

\textbf{ProbS}~\cite{zhou2010solving}  assigns a generic resource to each item $i_s$ held by a target user $u_t$. The resource is evenly split amongst the users directly connected to the item. Subsequently, each user evenly splits the portion of the resource received amongst the items connected to her. The final score $f^P_j$ of each item $i_j$ is given by the sum of the portions of resources that are assigned to it after the two steps, or, more formally: \vspace{-0.5cm}

\begin{equation}
  \vspace{-0.1cm}
	f^P_j = \sum_{l = 1}^{n}\sum_{s = 1}^{m} \frac{a_{l,j}a_{l,s}a_{t,s}}{k(u_l)k(i_s)} \hspace{1cm} j = 1, 2, \ldots, m
\label{eq:ProbS}
\end{equation}
where $k(u_l) = \sum_{j = 1}^{m} a_{l,j}$ is the number of items collected by the user $u_l$ and $k(i_s) = \sum_{j=1}^{n} a_{s,j}$ is the number of users interested in the item $i_s$. The set of $f^P_j$ values determines a ranking of contents concerning the interests of $u_t$.
ProbS tends to recommend items with the highest popularity. 

\textbf{HeatS}~\cite{zhou2010solving} uses rules opposite to those of ProbS. Each resource is first split amongst the items related to each user, and then amongst the users connected to each item. The score of the item $i_j$ for the target user $u_t$ is: \vspace{-0.5cm}

\begin{equation}
  \vspace{-0.2cm}
	f^H_j = \frac{1}{k(i_j)}\sum_{l = 1}^{n}\sum_{s = 1}^{m} \frac{a_{l,j}a_{l,s}a_{t,s}}{k(u_l)} \hspace{0.5cm} j = 1, 2, \ldots, m
\label{eq:HeatS}
\end{equation}

HeatS tends to recommend non-popular items. 

\textbf{Hybrid (ProbS + HeatS)}~\cite{liu2010improved} calculates a linear combination of ProbS and HeatS
using an \textit{hybridization parameter} $\lambda \in [0,1]$ such that by setting $\lambda = 0$ we obtain the pure HeatS, and with $\lambda = 1$ we get instead ProbS. The value of $\lambda$ may be difficult to select in real situations.

\textbf{PD and BHC}~\cite{Zhang2015Information} try to correct ProbS and HeatS. Preferential Diffusion (PD) divides the ProbS scores by the degree of the recommended item, with an exponent $\epsilon$ used as a parameter to control the normalization. Biased Heat Conduction (BHC) multiplies the HeatS score of each recommended item by its popularity, using an exponent $\gamma$ similar to $\epsilon$.  An optimal tuning of the parameters could be difficult to achieve in practice.

\vspace{-0.2cm}
\section{PLIERS}

PLIERS is inspired by ProbS and shares with it the same two  steps. In addition, PLIERS normalizes the value obtained by ProbS when comparing an item $i_j$ with one of the items of the target user, $i_s$, by multiplying the score by the cardinality of the intersection between the set of users connected to $i_j$ and the set of users connected to $i_s$, divided by $k(i_j)$ (i.e., the popularity of $i_j$). In this way, items with popularity similar to the popularity of the items of the target user, and which possibly share the same set of users, are preferred.
The score of the item $i_j$ is then: \vspace{-0.4cm}

\begin{equation}
  \vspace{-0.1cm}
  \small
  f^{PL}_j = \sum_{l = 1}^{n}\sum_{s = 1}^{m} \frac{a_{l,j} a_{l,s} a_{t,s}}{k(u_l) k(i_s)} \frac{\left | U_s \cap U_j \right |}{k(i_j)} \hspace{0.3cm} j = 1,\ldots,m
\label{eq:pliers_formula}
\end{equation}
where $U_j$ is the set of users connected to the item $i_j$ and $k(i_j)$ is the popularity degree of the item $i_j$. The normalization introduced in PLIERS favours items whose popularity (i.e.~number of connected users) is similar to that of the items already owned by the target user. All the procedures above can be equally applied to user-tag graphs, leading to the same considerations. 

\begin{table}[t]
  \caption{Datasets properties.}
  \vspace{-0.2cm}
  \label{tab:ranking_samples}
  \footnotesize
  \begin{tabular}{l  r  r  r  r  r }
    \toprule
    \textbf{Sample} & \textbf{Users} & \textbf{Tags} & \textbf{Links} & \textbf{$\overline{k}(T)$} & \textbf{$\overline{p}(T_U)$}          \\ \midrule
    MovieLens         & 5 K            & 17 K           & 105.6 K         & 6.14        & 52.75           \\
    Delicious         & 1.9 K          & 40.6 K         & 230.5 K         & 5.67        & 121.80          \\
    Twitter           & 5 K            & 194 K          & 508.5 K         & 2.62        & 74.24           \\
    \bottomrule
  \end{tabular}
  \vspace{-0.4cm}
\end{table}
\begin{table*}[t]
  \centering
  \vspace{-0.2cm}
\caption{Experimental results. Values in bold are related either to PLIERS or to the systems that outperform it.}
\label{tab:avg_pop_over}
\scriptsize
\begin{tabular}{l|rr|rrrrrrrrrr}
    \toprule
        {\bf }    & \multicolumn{2}{c|}{{\bf PLIERS}}              & \multicolumn{2}{c}{{\bf ProbS}}               & \multicolumn{2}{c}{{\bf HeatS}}               & \multicolumn{2}{c}{{\bf Hybrid}}              & \multicolumn{2}{c}{{\bf PD}}            & \multicolumn{2}{c}{{\bf BHC}}\\
        \midrule
& \multicolumn{1}{c}{\textit{V}} & \multicolumn{1}{c|}{\textit{O}} & \multicolumn{1}{c}{\textit{V}} & \multicolumn{1}{c}{\textit{O}} & \multicolumn{1}{c}{\textit{V}} & \multicolumn{1}{c}{\textit{O}} & \multicolumn{1}{c}{\textit{V}} & \multicolumn{1}{c}{\textit{O}}  & \multicolumn{1}{c}{\textit{V}} & \multicolumn{1}{c}{\textit{O}} & \multicolumn{1}{c}{\textit{V}} & \multicolumn{1}{c}{\textit{O}}\\
          \midrule
MovieLens & \textbf{41.90}                 & \textbf{0.118}                 & 80.34                 & 0.102                 & 50.50                 & 0.054                 & 50.82                 & 0.091                 & \textbf{41.54}          & 0.085         & 49.94           & 0.063\\
Delicious & \textbf{288.50}                & \textbf{0.090}                 & 422.87                & 0.085                 & \textbf{121.01}                & 0.007                 & 299.052               & 0.087          & \textbf{120.48}          & 0.026          & \textbf{181.08}         & 0.044\\
Twitter   & \textbf{91.01}                 & \textbf{0.017}                 & 560.36                & \textbf{0.021}                 & \textbf{73.22}                 & 0.001                 & 244.52                & \textbf{0.020}                 & \textbf{73.00} & 0.009 & \textbf{73.13} & 0.002\\
\bottomrule
  \end{tabular}
  \vspace{-0.4cm}
\end{table*}

\section{Experimental Results}
\label{sec:results}

We compared PLIERS with reference {\sc tbrs}s: HeatS, ProbS, Hybrid with $\lambda=0.5$; PD with $\epsilon=-0.85$ and BHC with $\gamma=0.8$ as in~\cite{Zhang2015Information}.  We used three benchmark datasets containing user-tag bipartite graphs. We assessed the accuracy of the obtained recommendations by calculating the level of personalization in terms of popularity of the recommended tags and the appropriateness of recommendations with respect to the users' interests. We performed also a link prediction task on the datasets~\cite{zhou2010solving, zhang2011tag, zhang2010personalized}. It consists in randomly removing a few links from the graph and to calculate the degree to which the recommendations coincide with the removed links. A good recommender system should be able to approximate the original graph, although  removing links changes the structure of the graph, and a complete reconstruction is not possible, particularly with sparse graphs.

\textbf{Datasets Description.}
\label{sec:datasets}
We used three bipartite user-tag graphs obtained from Twitter~\cite{dunbar2015structure}, MovieLens and Delicious~\cite{zhang2010personalized, zhou2010solving}. The graphs extracted from these datasets are very large (i.e., 1.6M users and 30.2M tags for Twitter, 1.9K users and 40.9K tags for Delicious, and 8.7K users and 39.2K tags for MovieLens). Due to memory constraints, we sampled portions of these graphs with maximum size of 5,000 users. Table~\ref{tab:ranking_samples} summarizes the characteristics of the obtained samples, where $U$, $T$, and $L$ are respectively the number of users, tags, and links. $\overline{k}(T)$ is the average tag degree in the graph and $\overline{p}(T_U)$ is the average popularity of the tags for the average user. From Table~\ref{tab:ranking_samples}, we can note that tags in Twitter are connected, on average, to fewer users than in the other datasets (i.e., $\overline{k}(T)$ is lower). This could lead to less accurate results in terms of link prediction.

\textbf{Metrics.}
\label{sec:metrics}
We defined an index $V$ (\emph{variance}), to calculate the average difference in terms of popularity between the recommended tags and those already owned by the users: \vspace{-0.5cm}

\begin{equation}
  \vspace{-0.2cm}
  V = \frac{1}{n} \sum_{l=1}^{n} \frac{1}{r_l} \sum_{q=1}^{r_l} \sqrt{(k(t_q) - p(T_{u_l}))^2}
\label{eq:var_mean}
\end{equation}
where $n$ is the number of users in the network, $r_l$ is the number of recommended tags for user $u_l$ and $p(T_{u_l}) = \frac{1}{z} \sum_{j=1}^{z} k(t_j)$ is the mean popularity of the tags originally linked to the user $u_l$ with $z$ the number of those tags. The \emph{overlap} $O$ measures the percentage of users connected to both the recommended tag and one of the tags of the target user, averaged for all the tags of the user and then for all the users. It gives us an idea of the potential interest for the users in the recommended tags.  It is defined as: \vspace{-0.5cm}

\begin{equation}
  \vspace{-0.2cm}
  O = \frac{1}{n} \sum_{l=1}^{n} \frac{1}{r_l} \sum_{q=1}^{r_l} \frac{1}{z} \prod_{k=1}^{z} J(U_{i_q}, U_{i_k})
\label{o_avg}
\end{equation}
where $U_{i_q}$ is the set of users connected to the item $i_q$ and $J(S_1, S_2)$ is the Jaccard's index, that measures the percentage of overlap between two generic sets $S_1$ and $S_2$. A good system should provide both a low $V$ and a high $O$.
\begin{figure*}[t]
  \centering
  \includegraphics[width=0.9\textwidth]{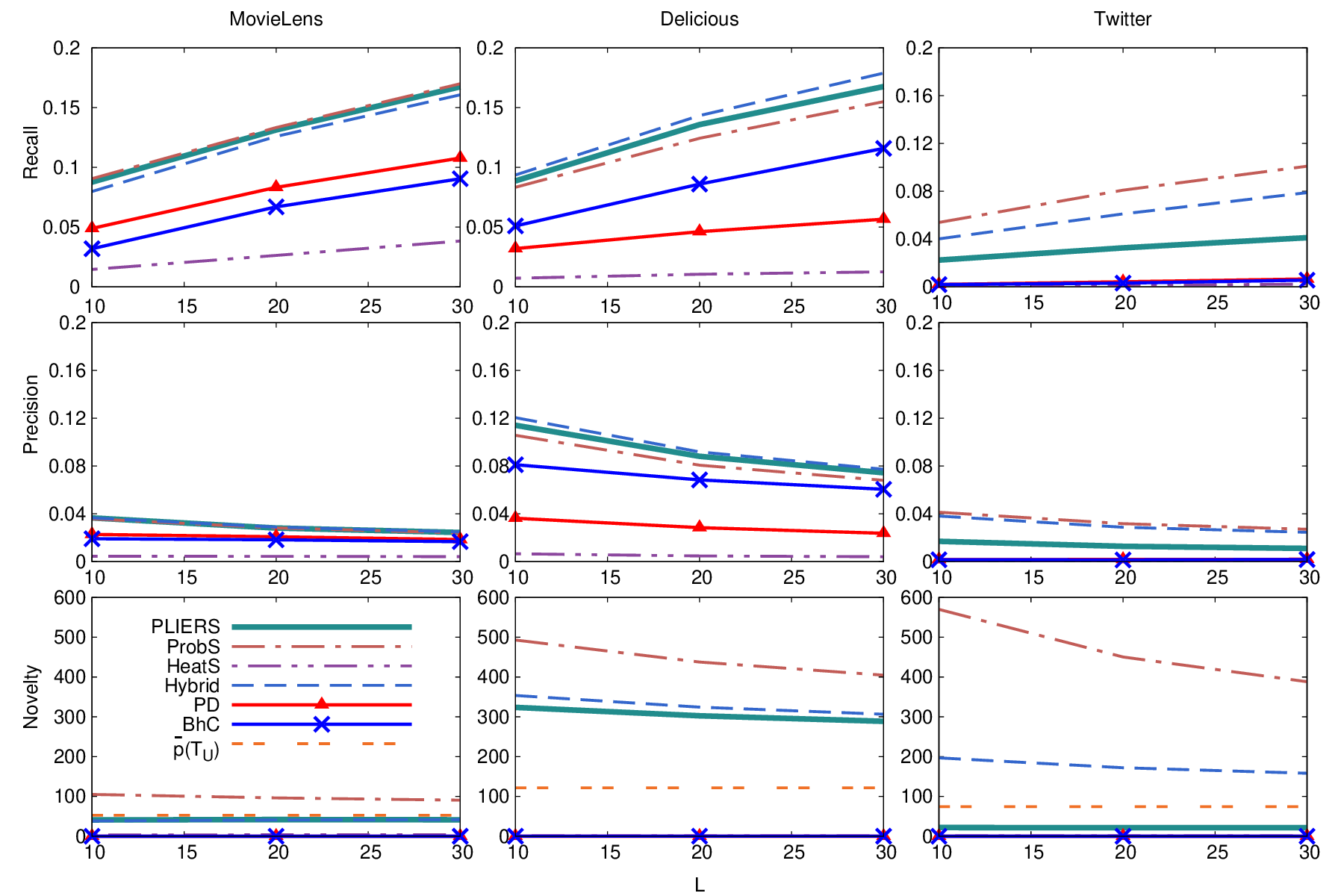}
  \vspace{-0.2cm}
  \caption{$R$, $P$ and $N$ with MovieLens, Delicious, and Twitter. $\overline{p}(T_U)$ is added to the novelty plots as a reference value.}
  \label{fig:linkp}
  \vspace{-0.4cm}
\end{figure*}
\\
For link prediction, we used three standard metrics. The \emph{recall} ($R$) index measures the number of recovered links within the first $L$ recommendations for each user divided by $L$. The \emph{precision} ($P$) measures the number of recovered links within the first $L$ recommendations divided by the total number of recovered links, for each user. The \emph{novelty} ($N$) index measures the capacity of a recommender system to generate novel and unexpected results, generally related to items with low popularity, quantified by measuring the average popularity of the first $L$ recommended items. A good system should have high $P$ and $R$, and low $N$.

\textbf{Results and Discussion.}
Table~\ref{tab:avg_pop_over} shows the values of $V$ and $O$ for the different datasets and {\sc tbrs}s. We highlight in bold the values better than those achieved by PLIERS. We note that PLIERS always yields the better trade-off.  As far as $V$ is concerned, PLIERS obtains values very close to the best results for two traces, and it always outperforms both ProbS and Hybrid.  It yields the best $O$, or very close to the best with Twitter. With Delicious, HeatS, PD, and BHC perform better than PLIERS in terms of $V$. Yet, with this trace, PLIERS supplies an overlap that largely outperforms those of the solutions yielding better $V$. These results tell that PLIERS is able to recommend tags whose popularity is comparable with those of the tags already owned by the users, and also of higher (or similar) relevance than the other solutions.
\\
Figure~\ref{fig:linkp} depicts the results of the link prediction task. As in~\cite{zhang2010solving}, we removed $10\%$ of the links. From the figure, we note that PLIERS again supplies the best trade-off. Its $R$ and $P$ are always very similar to the results of ProbS and Hybrid. In the case of Twitter, PLIERS' $P$ and $R$ are worse than those of ProbS and Hybrid, but in this case tags are connected, on average, to fewer users than in the other graphs and the removal of random links has a higher impact on the graph structure, having a negative impact on the recommendations. In this case, recommending tags with high popularity (as done by ProbS and Hybrid) is probably more effective. However, the level of personalization is clearly worse than the one obtained by PLIERS, as shown by the $V$ index. For the $N$ index, PLIERS is always better than ProbS and Hybrid, and reaches a value that is closer to the value of $\overline{p}(U_T)$. Hence, PLIERS is able to recommend tags of comparable popularity to that of the target user. 

\vspace{-0.4cm}
\section{Conclusions}
\label{sec:conclusion}
In this work, we proposed a new tag-based recommender systems called PLIERS that recommends tags or items with popularity as similar as possible to those already owned by the users. We compared PLIERS with other reference systems in the literature. The results indicate that PLIERS recommends tags with popularity closer to that of tags owned by the users than the other solutions. In case of link prediction, PLIERS performs very well, with results comparable to the other existing recommender systems in terms of precision and recall, but providing better novelty in the recommendations. 
\vspace{-0.4cm}
\section{Acknowledgment}

This work was partially funded by Registro.it within the Collective Awareness Participatory Platform research project (CAPP) and by EIT Digital within GameBus project.

\bibliographystyle{abbrv}
\vspace{-0.2cm}
\bibliography{SAC16_poster}




\end{document}